\documentclass{article}

\usepackage[utf8]{inputenc}

\title{Quantifying the Effects of the 2008 Recession using the Zillow Dataset}
\author{Arunav Gupta \and Lucas Nguyen \and Camille Dunning  \and Ka Ming Chan}
\date{
    Hal\i c\i o\u{g}lu Data Science Institute \\
    December 2019
}

\usepackage{graphicx}
\usepackage{amsmath}
\usepackage{subfigure}
\usepackage{layouts}
\DeclareMathOperator*{\argmin}{arg\,min}

\begin{document}

\graphicspath{{./images/}}

\maketitle
\section{Background of Crisis}
The 2008 Recession, also known as the “subprime mortgage crisis” was an economic recession that started in the US and quickly had global economic implications. From a period from late 2007 to mid-2009, 8.4 million Americans lost their jobs, 1 in every 54 homes filed for foreclosure, and the US GDP fell by 4.3 percent. 

Most economists agree that the crisis has its roots in the overuse of subprime mortgages, which were a type of security usually given to those who exhibit high financial risk and have low credit scores. Starting in the early 2000s, the housing industry was booming, and as houses steadily increased in value, more banks felt they could assume the risk of subprime mortgages and thus started giving them out en masse, even to those who would have traditionally been denied a loan. However, in 2007, the housing bubble burst and home values fell by almost 31.8\%. As demand for housing plateaued, home values dropped, and people who had taken out subprime loans found that they were no longer able to pay off the high interest payments associated with a subprime mortgage. As thousands defaulted on their mortgages, the lending institutions lost money, too. Compounding on the current problems, many of these institutions had traded mortgage-backed securities (MBSs), which were backed by these risky subprime loans, to other institutions seeking a profit when housing prices increased. When that didn’t happen, MBSs lost value as well, causing many banks like the Lehman Brothers to go bankrupt. The collapse of the real estate industry then caused banks and businesses to lose trust in each other, driving stock prices down. As publicly-traded companies then saw decreased valuations, businesses shut down, and unemployment skyrocketed. 

This “perfect storm” of financial disasters – subprime mortgages, housing value decline, mistrust in the stock market – hurt those at the bottom of the economic ladder the most, as without a job or a good credit rating to fall back on, they were unable to buy a home or provide for their families.

\subsection{Maps}
To visualise the recession effect on home values, we plotted the average ZHVI for each state per year, as well as the differences between each year. 

\begin{figure}[!h]
\centering
\subfigure[Figure A]{\label{fig:a}\includegraphics[width=.8\linewidth]{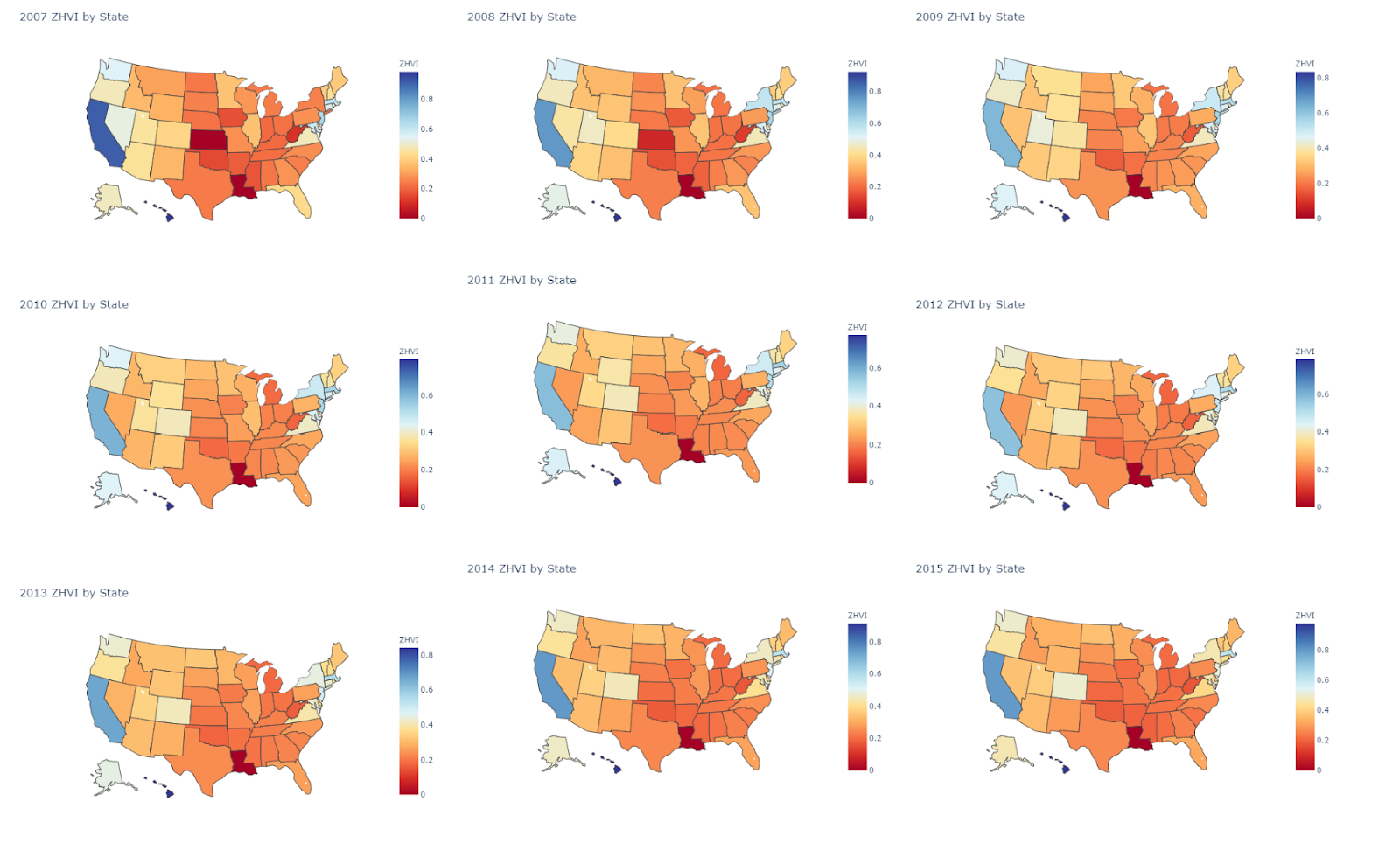}}
\subfigure[Figure B]{\label{fig:b}\includegraphics[width=.8\linewidth]{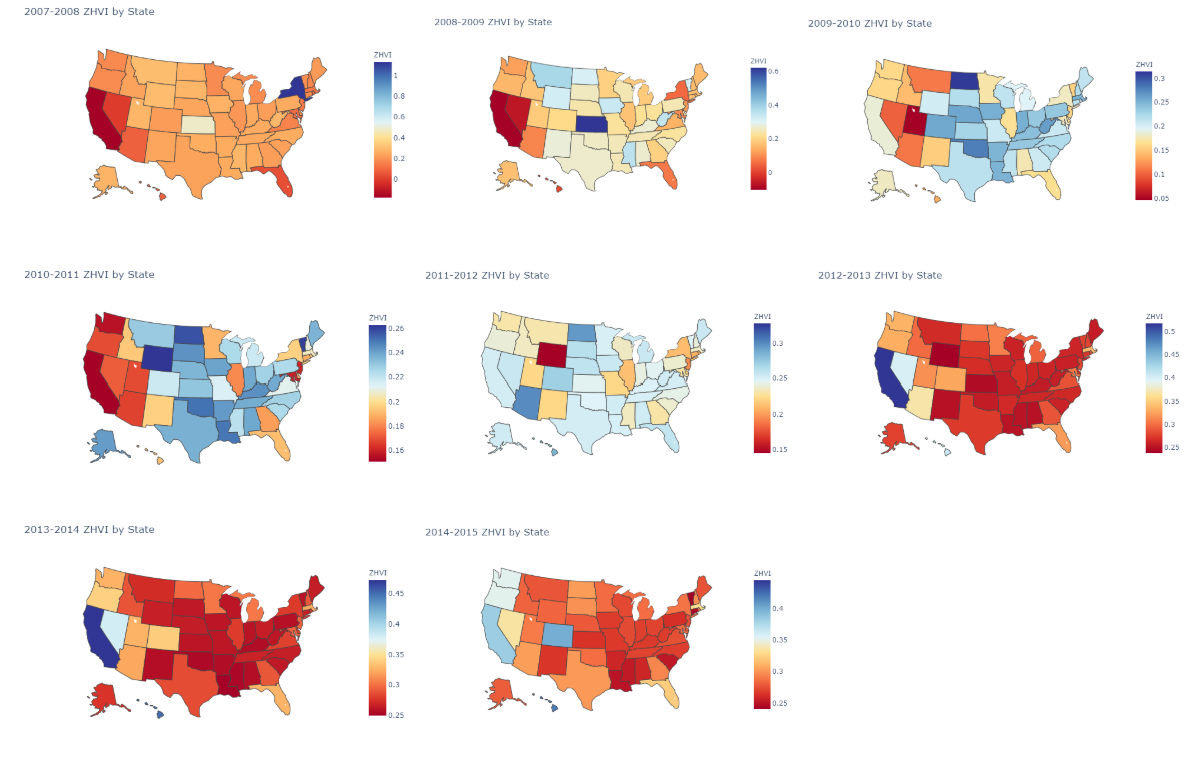}}
\caption{ZHVI, 2007-2015}
\end{figure}

The first map is normalized to a scale of 0 to 550,000 second map is normalized to a scale of -50,000 to +150,000. Overall we see a significant drop and then recovery reflected in the maps.

\section{Description of Dataset}
Our focus dataset was the ‘Zillow Economics Data’ found on Kaggle, which includes records of transactions, contracts, and specs about public properties throughout neighbourhoods, cities, statistical metropolitan areas, counties and states. The data are in a time series format starting from 1996 to around 2014. 

The observations in this dataset are derived from property listings and user behaviour on Zillow, and statistics were computed for metrics including the sales price, listing price, rent price. These statistics were acquired for properties per number of bedrooms, property type, property tier, and overall. Some other features include, but aren’t limited to, the number of days the property was listed on Zillow, the raw inventory, the percentage of annual increase or decrease in property value, sales turnover, and raw sales.

We decided to conduct our analyses on the property statistics across metropolitan areas. In our data, metropolitan areas are specified by a CBSA code. This code corresponds to a certain, CBSA, or core-based statistical area, which is characterised by one or more counties that are anchored by an urban centre. The Zillow Economics Data include both micropolitan and metropolitan CBSAs, where the urban centre of a micropolitan CBSA is between 10,000 and 50,000 and the metropolitan urban centre has greater than 50,000 people.

For reasons described in our section on the method of PCA, we decided not to use statistics for individual metrics as our dependent variables, but instead another provided measure, the Zillow Home Value Index (ZHVI). The ZHVI is based on the Zestimate home valuation model; it takes the median estimate in a geographic area on a given day. The median Zestimate is more sensible to use as it handles extreme values much better. Thus, we proceeded to use the ZHVI as the basis for our AUB and ARIMA implementations. There is an equivalent for rental spaces, the Zillow Rent Index (ZRI), but we didn’t run analyses for these properties.

\section{Methodologies}
\subsection{PCA}
The first methodology our group has performed for trial is called the Principal Component Analysis (PCA). The original dataset has approximately 95 columns of variables for us to build our analysis on. For our study, we would only like to extract a few useful features in the purpose of reducing the amount of excessive information. One way to reduce our high dimensional 95 columns of data into low dimensional representations of it is by using the method of PCA. In summary, the PCA selects a few important features from a sparse vector of data and compresses it by ignoring components which are not meaningful. Therefore, the data can be recovered and summarized as few dimensions as possible. The process can simply be described by figure 3 below:
\begin{figure}[!h]
\centering
\includegraphics[width=\textwidth]{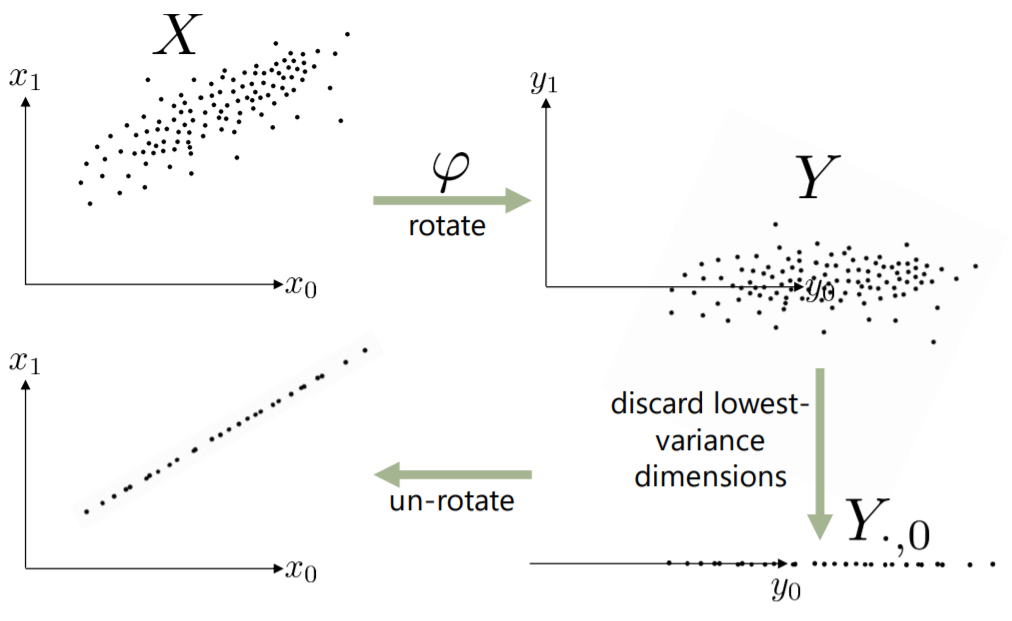}
\caption{Visualization of PCA}
\label{fig:pca_1}
\end{figure}
As shown in the process, the way we select the important features is by keeping the dimensions with the highest variance, and discarding the dimensions with the lowest variance. The highest variance dimensions maximize the amount of “randomness" that gets preserved in the compressed data. The method for determining the compressed data is simply by minimizing the Mean Square Error (MSE) between the original data and the new compressed line. 

We apply the PCA method onto our dataset and have reduced the 95 variables into 2. Here is a plot of the result in figure 4:
\begin{figure}[!h]
\centering
\includegraphics[width=\textwidth]{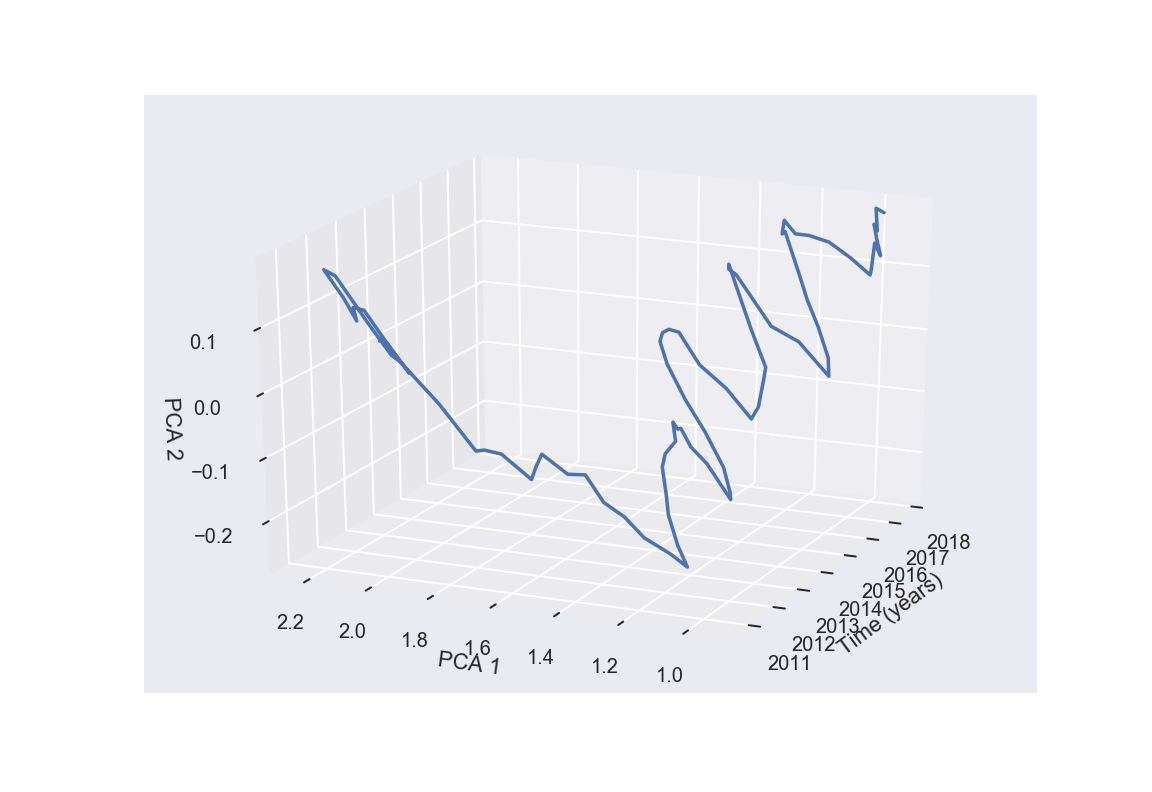}
\caption{Visualization of PCA on US nationwide data (n=3)}
\label{fig:pca_2}
\end{figure}

Unfortunately this graph does not provide much unique insights to us, as there is not much useful information from it that we can extract. The graph is messy, and not scalable because in the original dataset, most of the regions do not have complete data in them. In conclusion, the first methodology our group has tried, the PCA, does not reinforce us in finding any interesting correlation between our data and the recession.

\subsection{Area Under Baseline (AUB)}
The Area Under Baseline metric seeks to answer the question: “How much did the recession affect the ZHVI of a city?” To answer this question, we must look at the total impact the recession had on the city in question. There are multiple parts to the procedure: (a) transform the ZHVI data into a moving average (hereafter noted as $ZHVI_{MA}$), (b) find the “recession window” for the city, and (c) find the area between the  $ZHVI_{MA}$ trend and the baseline ZHVI across the recession window. 

Below, we’ll illustrate the procedure on Aberdeen, WA, which has a ZHVI trend graph that is very conducive to the process. Figure 5 shows Aberdeen’s  $ZHVI_{MA}$, computed with a window = 5. The red lines denote the recession window: the beginning of the window is the greatest local maximum of the $ZHVI_{MA}$ after Jan. 1, 2007. The end of the window is the point in time where the $ZHVI_{MA}$ intersects the value of $ZHVI_{MA}$ at the start point. The intuition here is that the recession is considered “recovered” once the ZHVI has returned to its pre-recession value. If the ZHVI never reaches its pre-recession value, the end date of the recession window is set at the last available value in the dataset. Figure 5 also shows the baseline (green line), which is defined as the $ZHVI_{MA}$ value at the start point of the recession window. Finally, we can compute the residual between the baseline and the $ZHVI_{MA}$ for each point in time between the start and end date of the recession window, then sum those values up to get the AUB for Aberdeen, WA.
\begin{figure}[!h]
\centering
\includegraphics[width=\textwidth]{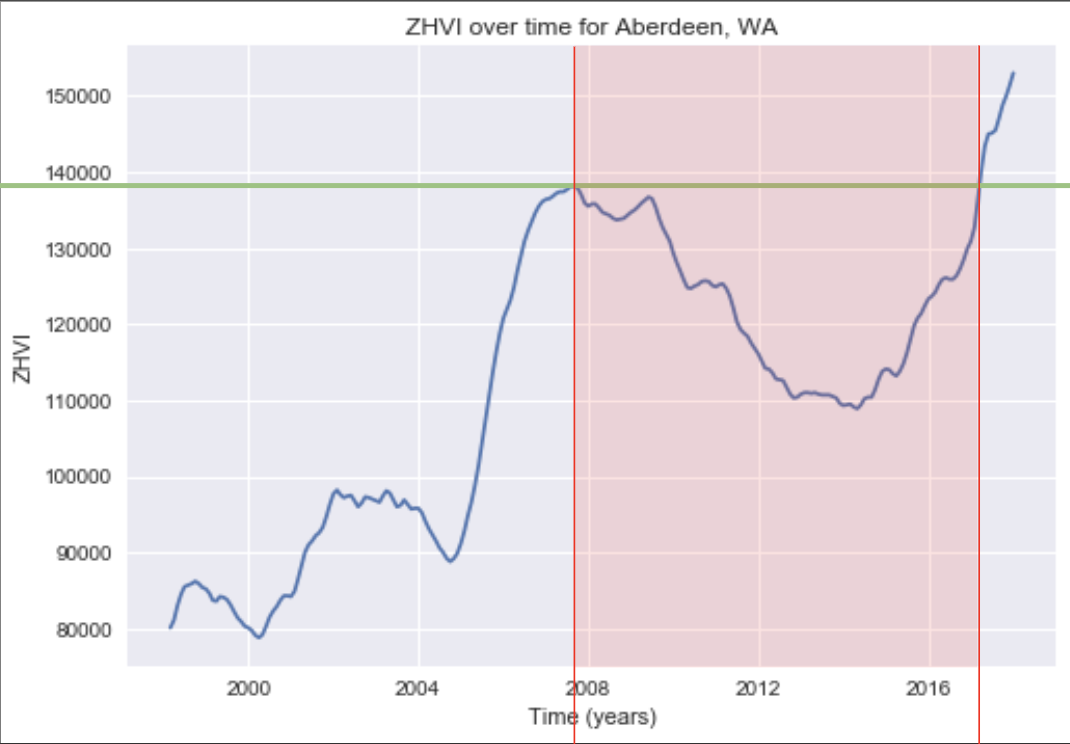}
\caption{$ZHVI_{MA}$ for Aberdeen, WA}
\label{fig:aberdeen}
\end{figure}

\subsubsection{Theory and Notation}
First we find the 5 month moving averages of the ZHVI values in the recession period
The 5 month Moving Average for month $i M_i$ is:\\
$M_{i}=\frac{1}{5}\sum_{k=0}^{4}y_{i-k}$ \\

$Y_i = $ ZHVI value at month i

We look for local maxes by finding dates where the moving average was positive before but negative after. If such local max exists, we find the local max with the largest ZHVI value: \\

$C=\left \{y_i :M_{i+1} < 0, M_{i-1} > 0\right \}$\\
$c = \max (C)$

The time that the local max occurs is declared the time window start. 

$a = i$

If it exists, we declare the time window end b to be the date of the first ZHVI value greater than the baseline. If not, we choose the date at the end of the recession period

$b = {\argmin}(c, y_z)$ \\
$y_z = $ ZHVI at end of recession period

Finally we find the area under the baseline by taking the sum of the differences between each ZHVI value in the time window and the baseline.

$\sum_{i=a}^{b}y_i-c$\\
$a = $ window start \\
$b = $ window end \\
$Y_i = $ ith ZHVI value in window \\
$c = $ baseline

\subsubsection{Results of Analysis}
Overall, the top 3 and bottom 3 AUB scores and their respective cities are listed in the tables below. Since a higher AUB score should be interpreted as an indicator of a higher recession impact, we will call the top 3 and bottom 3 cities “losers” and “gainers,” respectively:
\begin{table}[!h]
    \centering
    \begin{tabular}{c | c   ||   c | c }
    \textbf{Top "Loser" Cities} & \textbf{AUB} & \textbf{Top "Gainer" Cities} & \textbf{AUB}\\
    Key West, FL    & $1.638 \times 10^{11}$ & Mt. Vernon, IL & 17,080 \\
    Salinas, CA     & $1.541 \times 10^{11}$ & McAlester, OK  & 132,480 \\
    Carson City, NV & $1.331 \times 10^{11}$ & Norfolk, NE    & 287,040 \\
    \end{tabular}
    \caption{Results of AUB analysis}
    \label{tab:aub_results}
\end{table}
\subsection{ARIMA Model}
\subsubsection{What is Time Series Forecasting?}
There are many problems in predictive modeling that involve a time component. When we are making predictions about the outcome in the future, we are still treating all prior observations equally. In time series analysis, we have two different goals of either trying to understand and describe our time series data, or making predictions, or forecasting. Descriptive time series analysis can help with prediction, as it comprehends models to aid in identifying underlying causes, but it is not required and can be an investment.
Forecasting calls on models to fit historical data and using that information to predict future observations. 
"The purpose of time series analysis is generally twofold: to understand or model the stochastic mechanisms that gives rise to an observed series and to predict or forecast the future values of a series based on the history of that series."

To better understand time series analysis, we could decompose a time series into the following parts:
\begin{enumerate}
    \item \textbf{Level} - The average value in a series.
    \item \textbf{Trend} - The often linear increasing or decreasing behaviour of the series over time. - Optional, contingent on non-stationary or stationary time series
    \item \textbf{Seasonality} - Repeating patterns of cycles in behaviour over time. The ZHVI measure provided by the Kaggle data is already smoothed and seasonally adjusted.
    \item \textbf{Noise} - Variability in observations, unexplainable by model
\end{enumerate}

We can combine these components to provide an observed time series, and add them together to form our model:

$ y(t) = level_t + trend_t + seasonality_t + noise_t $ 

Time series data can also require ample scaling and cleaning to adjust for uneven frequency, time spacing, outliers, missing values, etc. The ZHVI data have been cleaned to supply the \texttt{metro\_data.csv} table.

The autoregressive integrated moving average (ARIMA) is a time-series fitted model designed to aid in descriptive analysis and forecasting of time-series data. ARIMA is often applied to data that show non-stationarity. As implied by the name, ARIMA has these key attributes:

\subsubsection{Autoregression}
ARIMA employs a simple autoregression (AR) model, in which observations from previous time steps are used as input to a regression equation to predict the next value. Formally, we can indicate an autoregressive model of order $p$ by the following:

$ AR(p) = X_t = c + \sum_{i = 1}^{p} \varphi_iX_t-i + \varepsilon_t $, 

where $c$ is a constant, $\varphi_1,...,\varphi_p$ are the parameters of the model, and $\varepsilon_t$ is noise. $p$, the order of the autoregressive model, represents the number of lags, or previous observed series values, to be included in the model.  

Since we are using regression with a neighborhood of terms, we can express this model equivalently with a backshift operator, $B$:

$ AR(p) = X_t = c + \sum_{i = 1}^{p} \varphi_iB^iX_t + \varepsilon_t $

A backshift, or lag, operator operates on an element of a time series to produce the previous element. Let us define an arbitrary time series $X = {X_1, X_2, ...}$. Then $ BX_t = X_{t-1}, \forall t > 1 $. The backshift operator can be raised to arbitrary integer powers so that $B^kX_t = X_{t-k}$.

\subsubsection{Integration}
The ARIMA model is \textbf{integrated}, meaning it uses a process known as differencing, in which observations at consecutive time steps are subtracted. This makes our non-stationary time series stationary, stabilising the mean by reducing trend. In the cases where seasonality will be reduced, the time series variance will also be stabilised. 

\subsubsection{Moving Average}
We express a moving average model of order $q$:

$ MA(q) = X_t = \mu + \varepsilon_t + \theta_1\varepsilon_{t-1} + ... + \theta_q\varepsilon_{t-q} = \mu + \varepsilon_t + \sum_{i = 1}^{q}\theta_i\varepsilon_{t-i}$, 

where $\mu$ is the mean of the series, $\theta_1,...\theta_q$ are the parameters of the model, and $\varepsilon_{t-1},...,\varepsilon_{t-q}$ are noise error terms. Write in terms of the backshift operator:

$ MA(q) = X_t = \mu + (1 + \theta_1B + ... + \theta_qB^q)\varepsilon_t $

Simply put, the average is represented here is represented as the central value of our set of numbers, but it's calculated for values of the dependent variable at different time intervals. The order $q$ denotes the size of the moving average window.

The \texttt{ARIMA()} model in Python accepts $p$, $d$, and $q$, where $d$ is the order of differencing. 

After importing all necessary libraries and table, we proceeded to the ARIMA analysis on a sample metropolitan area, San Diego. First, we produced visualis of the ZHVI trend and a correlogram. The correlogram, or autocorrelation plot, plots the sample correlations of the regression for each lag value. We want to choose a nonzero value of $p$ for our model such that the autocorrelation is high, so we can avoid overestimation or underestimation of true values for training of our forecasting model.
\begin{figure}[h]
\centering
\includegraphics[width=\textwidth]{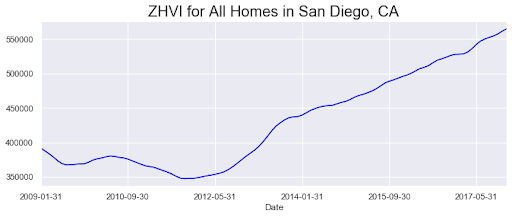}
\label{fig:ARIMA_1}
\end{figure}
\begin{figure}[h]
\centering
\includegraphics[width=\textwidth]{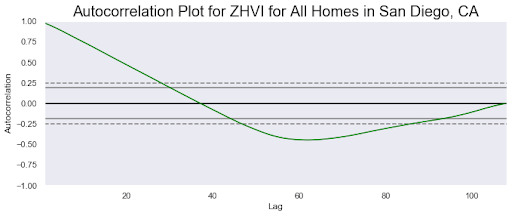}
\label{fig:ARIMA_2}
\end{figure}

As soon as we are done fitting the model, we will have a summary of the fit. We have also plotted the distributions for the residual errors, from which we can maybe capture some trend information. The density plot of the residual values show that they are Gaussian but not centred at zero. This is indicative of a bias in the prediction.
\begin{figure}[h]
\centering
\includegraphics[width=\textwidth]{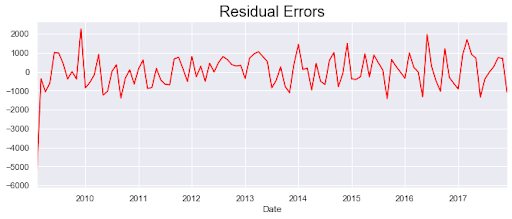}
\label{fig:ARIMA_3}
\end{figure}
\begin{figure}[!h]
\centering
\includegraphics[width=\textwidth]{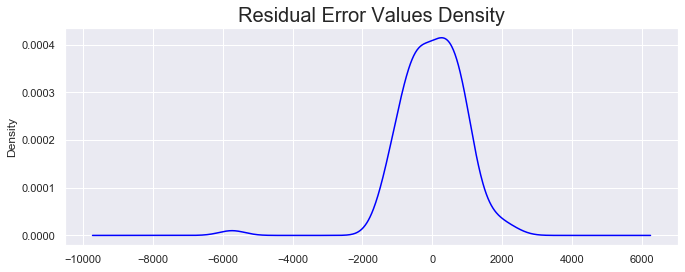}
\label{fig:ARIMA_4}
\end{figure}

\paragraph{We now test our model and produce a 95\% prediction interval for our forecasted results for the ZHVI in San Diego from 2017 to 2020. Note that there is a slight overlap with the in-sample and out-sample predictions.} 

\begin{figure}[!h]
\centering
\includegraphics[width=\textwidth]{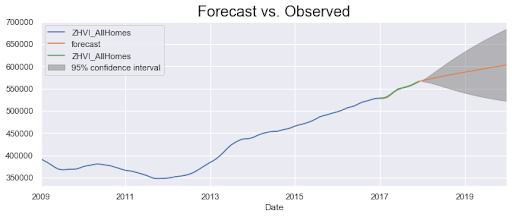}
\label{fig:ARIMA_5}
\end{figure}

\paragraph{Finally, we calculate the area of the confidence interval.}
\begin{figure}[!h]
\centering
\includegraphics[width=\textwidth]{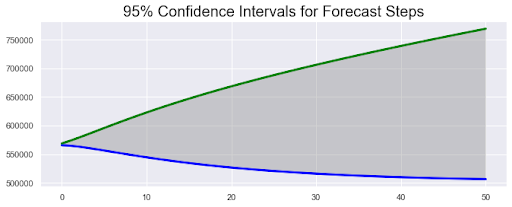}
\label{fig:ARIMA_6}
\end{figure}

\paragraph{The \textbf{smaller} the area of the 95\% confidence interval, the \textbf{less volatile the recovery over a longer period of time}, and the more certain we are that there will be a continuing increasing trend in the ZHVI. We proceeded to calculate the area of the 95\% prediction interval for each metropolitan area. The above shows the distribution of the normalised areas for across our observations.}

\begin{figure}[!h]
\centering
\includegraphics[width=\textwidth]{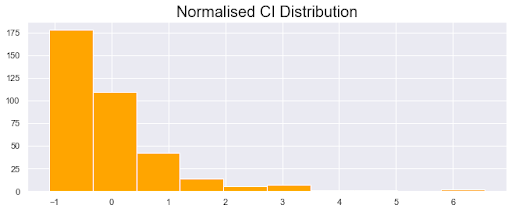}
\label{fig:ARIMA_7}
\end{figure}
\section{Results}
\subsection{Geographic Clustering}
By plotting the cities we received from each of the two working methodologies in the previous section (AUB and ARIMA), we can observe a couple patterns:
\begin{figure}[!h]
\centering
\includegraphics[width=\textwidth]{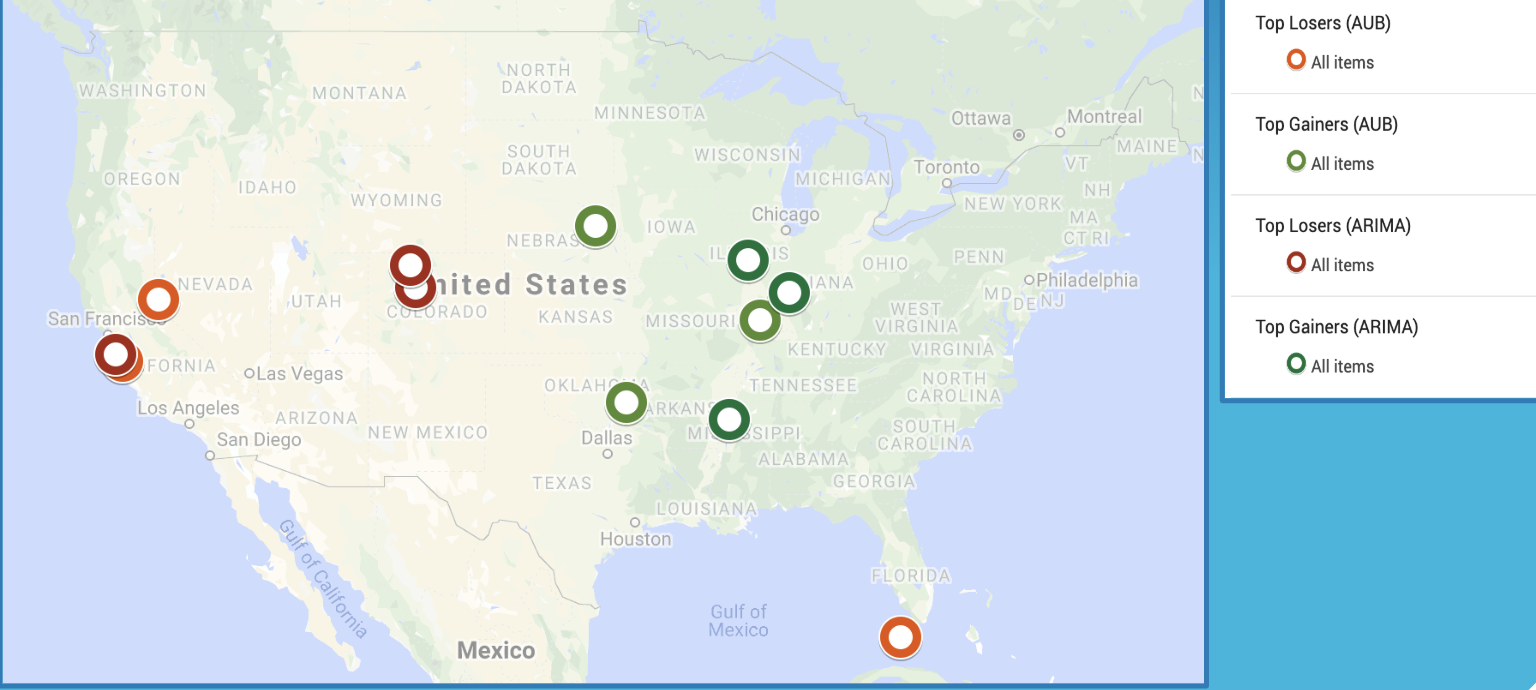}
\caption{Visualization of results from AUB and ARIMA}
\label{fig:map}
\end{figure}

First, we can notice that the gainer cities from both methods are primarily clustered in middle America/Great Plains, while many loser cities are located in the West. This defies conventional wisdom, which dictates that the West was not hit as hard as the Great Plains region due to its burgeoning technology industry. However, we can understand why Key West was marked as a loser, as it has a big tourism industry that was severely hit during the Recession, when less people could afford vacations and tourism.

\subsection{Metrics vs. Population}
We plot our AUB and ARIMA outputs against population for correlation. Figure 7 shows the two resulting graphs.

\begin{figure}[!h]
\centering     
\subfigure[Figure A]{\label{fig:a}\includegraphics[width=.4\linewidth]{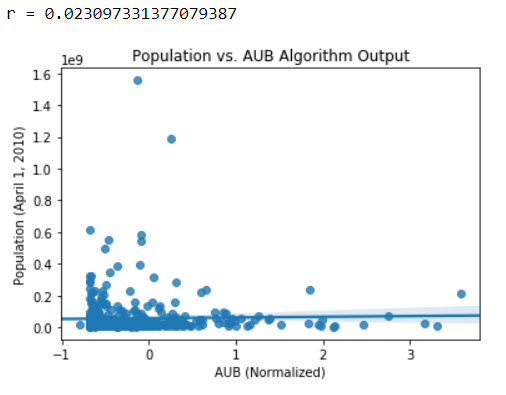}}
\subfigure[Figure B]{\label{fig:b}\includegraphics[width=.4\linewidth]{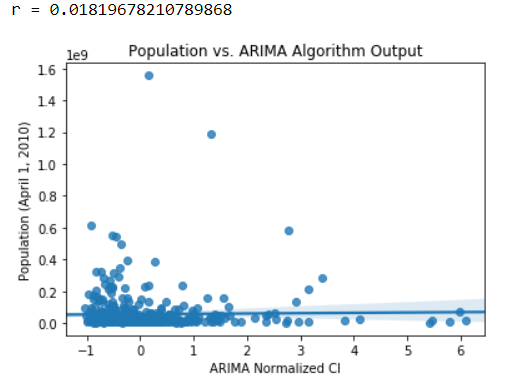}}
\caption{Regressing both methods against population statistics}
\end{figure}

Each blue dot represents a metro area. Since the r-value for both graphs are low, we conclude that there is little to no linear correlation between population and our algorithms. In the future we may consider transforming our data before testing for correlation.

\subsection{Metrics vs. Unemployment}
We have also plotted our AUB and ARIMA outputs against unemployment rates in hopes of finding any possible correlations. Figure 8 shows the two resulting graphs.

\begin{figure}[!h]
\centering     
\subfigure[Figure A]{\label{fig:a}\includegraphics[width=.4\linewidth]{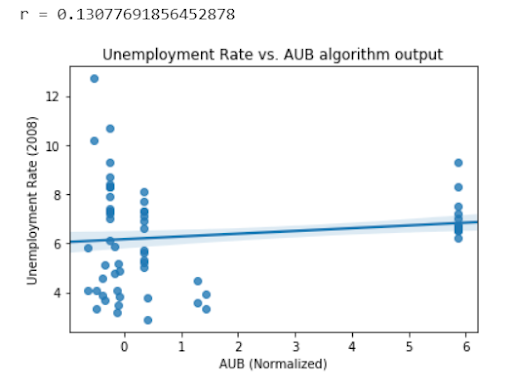}}
\subfigure[Figure B]{\label{fig:b}\includegraphics[width=.4\linewidth]{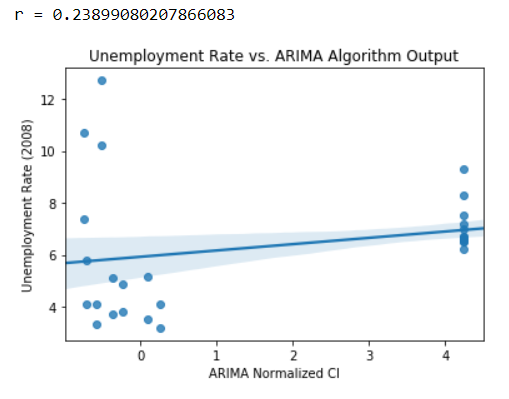}}
\caption{my caption}
\end{figure}

Each blue dot in the graphs represents a metro area. Judging from the regression lines of these two plots, there seems to be a correlation which reveals that the hardest-hit cities also had the biggest job losses. But as the R-squared values of them appear to be quite low, the correlations are not much secured. Therefore in conclusion, our attempts of plotting the population and the unemployment rates against our AUB and ARIMA outputs unfortunately do not show significant results.

\section{Conclusion}

Although we were able to come up with two unique metrics from determining the impact of the recession, neither of them seemed to agree with the universally-accepted metrics for determining recession impact – unemployment rate and city size. Our reasons for reaching this conclusion are because the AUB and ARIMA scores do not correlate very well with population and unemployment rate intuitively. This could be for a variety of reasons. The most prevalent is the notion that perhaps housing data such as the ZHVI is not the best indicator of a recession, after all. Another issue could be noise in the data. In many cases, the average ZHVI of a given metropolis was much larger than that of any other one, a feature of the dataset we could have corrected for by normalizing each metro’s data before applying a moving average.

\subsection{Future Steps}
Moving forward, some steps we may take are as follows: 
\begin{enumerate}
    \item An investigation of the spillover effect in our data.
    \begin{enumerate}
        \item According to a CityLab article, the spillover effect occurs when economic event that occurs in one city or metropolitan area will again occur in an adjacent city or metropolitan area. The study in the article pointed out that this was evident in a number of metros Chicago, New York, and Hartford, while Washington, D.C., Austin, and Providence were able to ride out the recession on their own. Essentially, we will want to check if our data exhibit possible relationships between ZHVI trends in one city and those of its neighbouring cities.
    \end{enumerate}
    \item A new algorithm that combines the AUB method and ARIMA forecasting. This may involve tuning hyperparameters of our ARIMA model to get more detailed predictive curves/trends on which we can then apply the Area Under Baseline.
    \begin{enumerate}
        \item We can visualise the median algorithm outputs per each state on a map, similar to those shown toward the beginning of the paper.
        \item With enough tweaking, we could use this algorithm to predict the effect of the 2020 recession on ZHVI values. This will involve more research on how to work with non-stationary time data, and may possibly yield the use of another more flexible time-series module.
    \end{enumerate}
    \item Comparing methodologies of other researchers in acquiring results for recession effects on housing metrics, and seeing if the results we attained stack up. This will provide us more insight as to what machine-learning approach we should employ with the sort of data we are given. 

\end{enumerate}


\end{document}